\documentclass[a4paper,12pt,english]{iopart}
\usepackage[T1]{fontenc}
\usepackage[utf8]{inputenc}

\usepackage{graphicx}
\usepackage{iopams}
\usepackage{setstack}
\usepackage{multirow}
\usepackage{cite}
\usepackage{setspace}
\usepackage{bm}
\usepackage{tikz}
\usepackage{type1cm}

\usepackage{babel}

\definecolor{ourBlue}{HTML}{0000FF}
\definecolor{ourOrange}{HTML}{FFA400}
\definecolor{ourGreen}{HTML}{008B00}
\definecolor{ourMagenta}{HTML}{FF34A7}
\definecolor{ourPurple}{HTML}{632280}
\definecolor{ourBrown}{HTML}{5E3100}

\definecolor{spin1}{RGB}{0,0,255}
\definecolor{spin0}{RGB}{255,165,10}
\definecolor{isc}{RGB}{0,139,0}
\definecolor{ir}{RGB}{255,52,167}

\definecolor{colourPi}{RGB}{0,139,0}
\definecolor{colourSig}{RGB}{255,52,167}

\newcommand*\circled[1]{\tikz[baseline=(char.base)]{
	\node[shape=circle,draw,inner sep=1pt] (char) {#1};}}

\newcommand{\tripAtwo}{{}^3\mathrm{A}_2}
\newcommand{\tripE}{{}^3\mathrm{E}}
\newcommand{\singAone}{{}^1\mathrm{A}_1}
\newcommand{\singE}{{}^1\mathrm{E}}

\newcommand{\doubE}{{}^2\mathrm{E}}

\newcommand{\Aone}{\mathrm{A}_1}
\newcommand{\Atwo}{\mathrm{A}_2}

\newcommand{\Ex}{\mathrm{E}_\mathrm{X}}
\newcommand{\Ey}{\mathrm{E}_\mathrm{Y}}

\newcommand{\A}{\mathrm{A}}
\newcommand{\E}{\mathrm{E}}

\renewcommand{\e}{\mathrm{e}}
\newcommand{\ai}{\mathrm{a}_1}

\newcommand{\NVvis}{\mathrm{NV}_{\mathrm{vis}}^-}
\newcommand{\NVir}{\mathrm{NV}_{\mathrm{IR}}^-}

\newcommand{\NV}{\mathrm{NV}}

\newcommand{\Cs}{\mathrm{C}_\mathrm{s}}
\newcommand{\Cthreev}{\mathrm{C}_\mathrm{3v}}

\newcommand{\IR}{_\mathrm{IR}}
\newcommand{\vis}{_\mathrm{vis}}

\begin{document}

\title{Singlet levels of the NV$^{-}$ centre in diamond}

\author{L J Rogers$^{1,2,3}$, M W Doherty$^1$, M S J Barson$^1$, S Onoda$^4$, T Ohshima$^4$
and N B Manson$^1$}
\address{$^1$ Laser Physics Centre, Research School of Physics and Engineering, Australian National University, Canberra,
ACT 0200, Australia}
\address{$^2$ School of Science and Mathematics, Avondale College of Higher Education, Cooranbong, NSW 2265, Australia }
\address{$^3$ Institut für Quantenoptik, Universität Ulm, Ulm, Germany}
\address{$^4$ Semiconductor Analysis and Radiation Effects Group, Japan Atomic Energy Agency, 1233 Watanuki, Takasaki, Gunma,370-1292, Japan}
\ead{lachlan.j.rogers@quantum.diamonds}

\begin{abstract}
The characteristic transition of the $\mbox{NV}^{-}$ centre at 637\,nm is
between $\tripAtwo$ and $\tripE$ triplet states.
There are also intermediate $\singAone$ and $\singE$ singlet states, and the
infrared transition at 1042\,nm between these singlets is studied here using
uniaxial stress.
The stress shift and splitting parameters are determined, and the physical interaction
giving rise to the parameters is considered within the accepted electronic
model of the centre.
It is established that this interaction for the infrared transition is due to a
modification of electron-electron Coulomb repulsion interaction.
This is in contrast to the visible 637\,nm transition where shifts and
splittings arise from modification to the one-electron Coulomb interaction.
It is also established that a dynamic Jahn-Teller interaction is associated
with the singlet $\singE$ state, which gives rise to a vibronic level
115\,cm$^{-1}$ above the $\singE$ electronic state.
Arguments associated with this level are used to provide experimental
confirmation that the  $\singAone$ is the upper singlet level and $\singE$ is the lower
singlet level.

\end{abstract}


\vspace{2pc}


\submitto{New Journal of Physics}

\maketitle


\section{Introduction}

\begin{figure}
\hfill
\includegraphics[width=134.6mm]{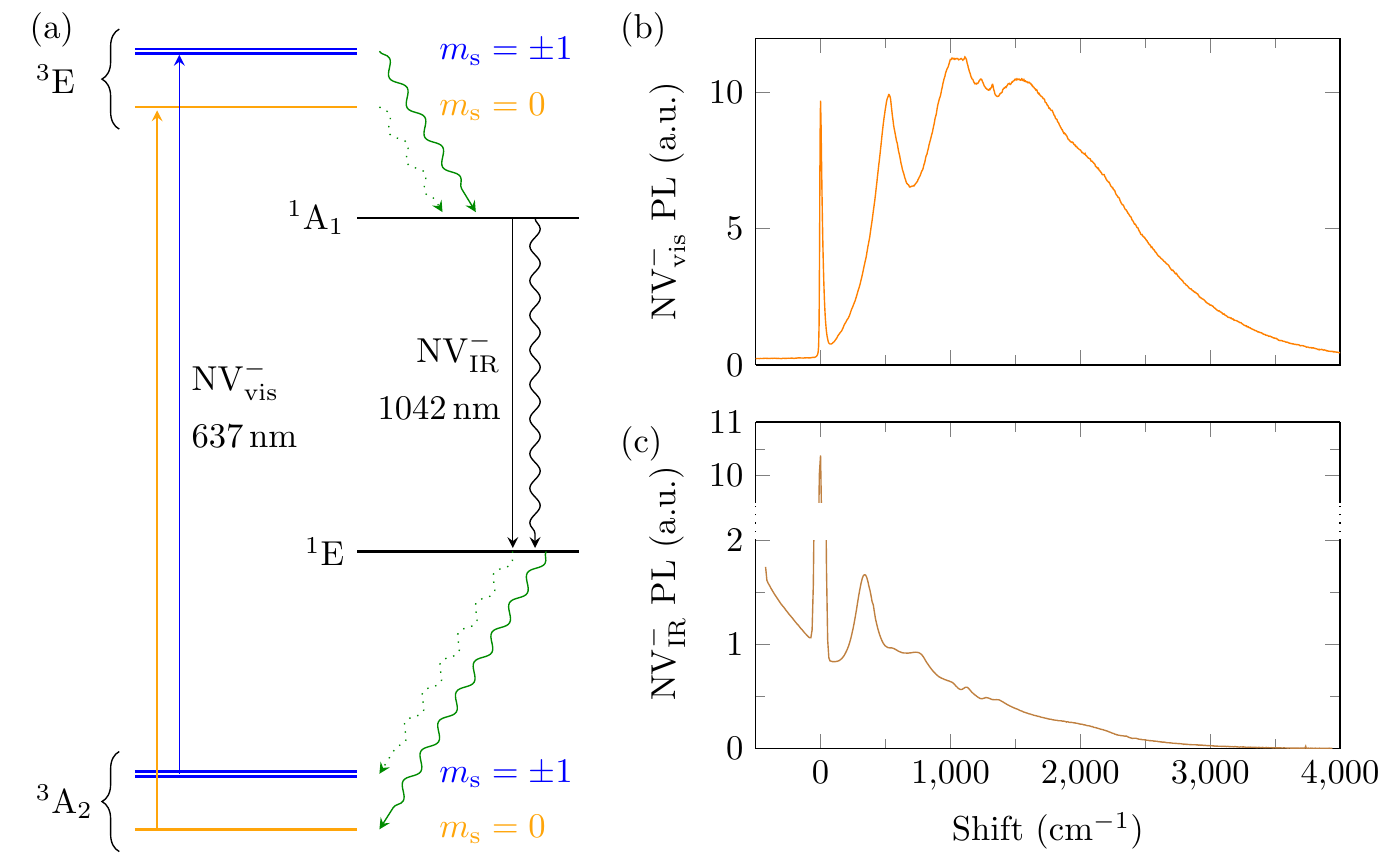}
\caption{
	Electronic energy level scheme and fluorescence bands for the NV${}^-$ transitions.
	(a)	The primary transition between triplet ground and excited states is
	predominantly spin conserving.
	Decay via the intermediate singlets gives rise to spin polarisation by
	preferentially switching spin from $m_\mathrm{s}=\pm1$ to $m_\mathrm{s}=0$.
	(b) The low temperature (10K) emission spectrum for the visible transition
	$\NVvis$.
	Emission was excited with 100 mW laser at 532 nm.
	(c) The $\NVir$ infrared band lies on the tail of the visible emission and
	has an integrated area of 1 $\pm$ 0.2 x 10$^{-3}$ compared to that of the
	visible band.
	It is understood that the weakness of this fluorescence band
	is due to strongly competing non-radiative decay between the singlets
	illustrated by the wavy arrow in (a) \cite{rogers2010how}.
	}
	\label{fig:spectra}
\end{figure}

The negatively charged nitrogen vacancy centre in diamond ($\mbox{NV}^{-}$)
\cite{doherty2013nitrogen-vacancy} exhibits optically induced spin
polarisation.
This property underpins many exciting applications of the $\mbox{NV}^{-}$
centre in fields such as magnetic sensing \cite{chernobrod2005spin,
degen2008scanning, taylor2008high-sensitivity, maze2008nanoscale,
balasubramanian2008nanoscale, hall2009sensing, cole2009scanning,
hall2010ultra-sensitive}, biological imaging \cite{fu2007characterization,
chang2008mass, tisler2009fluorescence}, and quantum information processing
\cite{gaebel2006room-temperature, dutt2007quantum, togan2010quantum,
neumann2010quantum, fuchs2008excited-state}.
The principle zero-phonon line (ZPL) associated with the centre is at
$637\,\mbox{nm}$ ($1.945\,\mbox{eV}$, 15687\,cm$^{-1}$) and is found by uniaxial
stress to involve a transition between a ground state of $\A$ symmetry and an
excited state of $\E$ symmetry at a trigonal site \cite{davies1976optical}.
Here we label this transition $\NVvis$ since it is in the visible spectrum, and
its fluorescence band is shown in Figure \ref{fig:spectra}.
The ground and excited states are spin triplets \cite{loubser1977optical,
loubser1978electron, reddy1987two-laser, oort1988optically, redman1991spin} and
optical excitation of this transition results in the spin being polarised into
$m_\mathrm{s}=0$, although this does not arise from direct optical cycling as
the optical transitions are spin-conserving \cite{manson2006nitrogen-vacancy}.
When the triplet system is excited there is also relaxation via intermediate
singlets and this decay causes the spin polarisation. A weak emission band in
the infrared (Figure \ref{fig:spectra}) with a ZPL at 1042\,nm
($1.19\,\mbox{eV}$, 9597 cm$^{-1}$) is associated with decay between these two
singlet levels  \cite{rogers2008infrared}.

A study of this emission (which we call $\NVir$) provides an opportunity to
better understand the electronic levels in this important decay channel.
Uniaxial stress is the experimental technique of choice.
A previous uniaxial stress study has shown that the $\NVir$ transition is
between levels of $\A$ and $\E$ symmetry \cite{rogers2008infrared}, and this
symmetry assignment is not in question.
However, in that study the specific transitions were not correctly identified
and this led to an inaccuracy of the stated stress parameters.
Here the transitions are unambiguously identified and correct stress parameters
are determined.
In addition we experimentally resolve the long-standing contention regarding
the order of the singlets \cite{goss1996twelve-line, manson2007issues,
gali2008ab, rogers2008infrared, delaney2010spin-polarization} and establish the
$\singE$ to be the lower singlet.
The magnitudes of the stress parameters are considered within the current
electronic model of the centre.
It is concluded that the interaction giving rise to the shift and splitting of
the infrared ZPL is different from that giving rise to the shifts and
splittings of the $\NVvis$ ZPL and $\mbox{NV}^{0}$ ZPL.

\section{Uniaxial stress theory}

The theory for uniaxial stress applied to an $\A\! \leftrightarrow\! \E$
transition at a site of trigonal symmetry in a cubic crystal has been given on
several occasions \cite{kaplyanskii1964noncubic, kaplyanskii1964computation,
hughes1967uniaxial, crowther1967phonon} and has been developed by Davies and
Hamer \cite{davies1976optical, davies1979dynamic} for the case of the NV
centre.
The elements of the stress tensor $s_{ij}$  as applied to the cubic crystal can be
expressed in terms of the irreducible representations appropriate for the
trigonal site symmetry, and the stress perturbation at the NV site is given by
\begin{eqnarray}
H^{s} &= \bm{A}_1(s_{xx}+s_{yy}+s_{zz})+
			\bm{A}_1^{'}(s_{yz}+s_{zx}+s_{xy})\nonumber \\
&\quad +\bm{E}_\mathbf{X}(s_{xx}+s_{yy}-2s_{zz})+
			\bm{E}_\mathbf{Y}\sqrt{3}(s_{xx}-s_{yy})\nonumber \\
&\quad +\bm{E}_\mathbf{X}^{'}(s_{yz}+s_{zx}-2s_{xy})+
			\bm{E}_\mathbf{Y}^{'}\sqrt{3}(s_{yz}-s_{zx})
\end{eqnarray}
where $\bm{A}_{1}$, $\bm{A}_{1}^{'}$ are symmetry adapted electronic operators
transforming as A$_1$ irreducible representations and $\bm{E}_\mathrm{X}$,
$\bm{E}_\mathrm{Y}$, $\bm{E}_\mathrm{X}^{'}$, $\bm{E}_\mathrm{Y}^{'}$ are operators
transforming as components of E irreducible representations
\cite{hughes1967uniaxial, davies1976optical}.
The stress $s_{ij}$ is given in terms of the lattice co-ordinates.
The effects of this interaction on an $\A\! \leftrightarrow\! \E$
transition have been described by Davies and Hamer
\cite{davies1976optical} in terms of the following reduced matrix elements
\begin{eqnarray}
A\!\mathit{1} &=& \langle \E\|\bm{A}_{1}\| \E \rangle -
			\langle \A \|\bm{A}_{1}\| \A \rangle, \\
2A\!\mathit{2} &=& \langle \E \|\bm{A}_{1}^{'}\| \E \rangle  -
			\langle \A \|\bm{A}_{1}^{'}\| \A \rangle,\\
\sqrt2 B &=& \langle \E \|\bm{E}\| \E \rangle, \\
\sqrt2 C &=& \langle \E \|\bm{E}^{'}\| \E \rangle  .
\end{eqnarray}
For stress applied
along  $\langle001\rangle$,  $\langle111\rangle$ and  $\langle110\rangle$
crystallographic directions, the resultant relative strength and polarisation of the transitions have been given
in previous publications \cite{kaplyanskii1964noncubic,
kaplyanskii1964computation, hughes1967uniaxial, crowther1967phonon,
davies1976optical} and are summarised in Table 1.
Since the NV centre is now known to involve both $\Aone\! \leftrightarrow\! \E$
and $\Atwo\! \leftrightarrow\! \E$ transitions as indicated in Figure 1 the
selection rules for both cases have been included in Table 1.
Stress along $\langle001\rangle$,  $\langle111\rangle$ or $\langle110\rangle$
directions is always in a reflection plane  or at right angles to a reflection
plane,  and consequently the site symmetry is always lowered to $\Cs$.
Therefore, for every case the  $\Gamma_1$ or $\Gamma_2$ irreducible
representations for $\Cs$ are included in the table.

\begin{table}
\caption{
	Summary of shifts, splittings and polarisation for stress applied along
	several crystallographic directions.
	The values are from reference \cite{davies1976optical} although here the
	values are normalised to an intensity of 8/3 at zero stress (each of the 4
	orientations contributing a relative oscillator strength of 2).
	Intensities are given for $\pi$ (electric field vector parallel to
	stress) and $\sigma$ (perpendicular) polarisations.
	The selection rules were given for $\Aone\!\leftrightarrow\!\E$
	transitions \cite{davies1976optical} and are extended here to also cover
	$\Atwo\!\leftrightarrow\!\E$ transitions.
	The change results in an interchange of X and Y and change of sign of $B$
	and $C$.}
\small
\onehalfspacing
\begin{tabular}{llcc|ccc|ccc}
\hline
 &&&E state&\multicolumn{3}{|c|}{$\Atwo\!\leftrightarrow\!\E$}&\multicolumn{3}{|c}{$\Aone\!\leftrightarrow\!\E$} \tabularnewline \hline
\hspace{3mm}Stress&\hspace{2mm}Orientation&Sym&Energy&$\pi$&\multicolumn{2}{c|}{$\sigma$}&$\pi$&\multicolumn{2}{c}{$\sigma$} \tabularnewline\hline
\multirow{4}{*}{\hspace{-3mm}\includegraphics[scale=0.8]{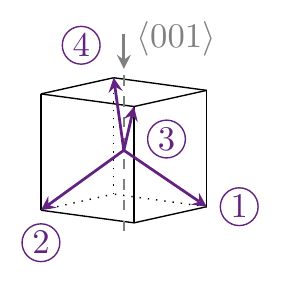}}\hspace{-3mm} & \multirow{4}{*}{$\begin{array}{l}\circled{1}\\[-4pt]\circled{2}\\[-4pt]\circled{3}\\[-4pt]\circled{4}\end{array}  54^\circ (\mathrm{XZ})$} & & & & & & & & \\
 & &$\Ex\,(\Gamma_1)$&$A\!\mathit{1}+2B$ &0&\multicolumn{2}{c|}{2}&$\frac{8}{3}$&\multicolumn{2}{c}{$\frac{2}{3}$}\tabularnewline
 & &$\Ey\,(\Gamma_2)$& $A\!\mathit{1}-2B$ &$\frac{8}{3}$&\multicolumn{2}{c|}{$\frac{2}{3}$}&0&\multicolumn{2}{c}{2}\tabularnewline
 & & & & & & & &  &\\\hline

\multirow{5}{*}{\hspace{-3mm}\includegraphics[scale=0.8]{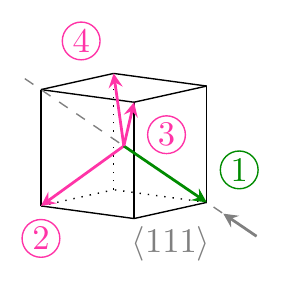}}\hspace{-3mm} & $ \begin{array}{c} \circled{1} \end{array}  0^\circ $  & $\Ex,\Ey$ & $A\!\mathit{1}+ 2A\!\mathit{2}$ &0&\multicolumn{2}{c|}{1}&0&\multicolumn{2}{c}{1}\tabularnewline
 & \multirow{4}{*}{$\begin{array}{l}\circled{2}\\[-4pt]\circled{3}\\[-4pt]\circled{4}\end{array} 70^\circ (\mathrm{XZ})$} & & & & & & &  &\\
&& $\Ex\,(\Gamma_1)$& $A\!\mathit{1}-\frac{2}{3}A\!\mathit{2} + \frac{4}{3}$C&0&\multicolumn{2}{c|}{$\frac{3}{2}$}&$\frac{8}{3}$&\multicolumn{2}{c}{$\frac{1}{6}$}\tabularnewline
&& $\Ey\,(\Gamma_2)$& $A\!\mathit{1}-\frac{2}{3}A\!\mathit{2}-\frac{4}{3}C$ & $\frac{8}{3}$&\multicolumn{2}{c|}{$\frac{1}{6}$}&0&\multicolumn{2}{c}{$\frac{3}{2}$}\tabularnewline
 & & & & & & & &  &\\\hline
 & & & & & $\sigma_{110}$ & $\sigma_{001}$ & &$\sigma_{110}$&$\sigma_{001}$ \\
\multirow{5}{*}{\hspace{-3mm}\includegraphics[scale=0.8]{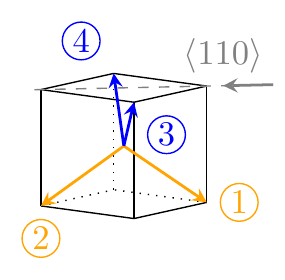}}\hspace{-3mm} &\multirow{2}{*}{$\begin{array}{l}\circled{1}\\[-4pt]\circled{2}\end{array} 36^\circ (\mathrm{XZ})$ }& $\Ex\,(\Gamma_1)$& $A\!\mathit{1}+A\!\mathit{2}-B+C$ &0&2&0&$\frac{2}{3}$&0&$\frac{4}{3}$\tabularnewline
&& $\Ey\,(\Gamma_2)$& $A\!\mathit{1}+A\!\mathit{2}+B-C$ &$\frac{2}{3}$&0&$\frac{4}{3}$&0&2&0\tabularnewline
 & & & & & & & &  &\\
&\multirow{2}{*}{$\begin{array}{l}\circled{3}\\[-4pt]\circled{4}\end{array} 90^\circ (\mathrm{YZ}) $}& $\Ex\,(\Gamma_1)$& $A\!\mathit{1}-A\!\mathit{2}-B-C$ &2&0&0&0&$\frac{2}{3}$&$\frac{4}{3}$\tabularnewline
&& $\Ey\,(\Gamma_2)$& $A\!\mathit{1}-A\!\mathit{2}+B+C$ &0&$\frac{2}{3}$&$\frac{4}{3}$&2&0&0\tabularnewline
\hline
\end{tabular}
\end{table}

\section{Experimental details}

Diamond cubes with dimensions $2\times2 \times 2\,\mathrm{mm}$ were used.  They
have nitrogen concentrations of $\sim$\,100\,ppm and were irradiated and
annealed to give $\mathrm{NV}^{-}$ concentrations of $\sim$\,5\,ppm.
The cubes  had either $\langle110\rangle$, $\langle1\bar{1} 0\rangle$  and
$\langle001\rangle$  faces or $\langle111\rangle$, $\langle1\bar{1}0\rangle$
and $\langle11\bar{2}\rangle$ faces.
These were used for application of stress along $\langle001\rangle$,
$\langle111\rangle$ and $\langle110\rangle$ directions by means of a pneumatic
driven rod.
The samples were within a cryostat and could be cooled to liquid helium or
liquid nitrogen temperatures as required.

For the majority of the work the emission was excited  by a laser at a wavelength of 532\,nm within the vibrational sideband of the $\tripAtwo\!
\rightarrow\! \tripE$ absorption transition.
The emission at right angles was dispersed by a monochromator and detected by a
photomultiplier (for $\NVvis$) or a cooled germanium detector (for $\NVir$).
A tunable dye laser at the wavelength of the visible ZPL was used
for selective excitation techniques to assist with the assignments of the
$\NVir$ spectra.

\section{Results}

\subsection{Uniaxial stress measurements along $\langle001\rangle$ and $\langle111\rangle$}

Although both the visible \cite{davies1976optical} and infrared
\cite{rogers2008infrared} transitions involve an $\A\! \leftrightarrow\! \E$
transition at a site of trigonal symmetry, $\NVvis$ involves an $\Atwo$ state
whereas $\NVir$ involves an $\Aone$.
In addition, the $\E$ state is the upper level for $\NVvis$ but for $\NVir$ the $\E$ is the
lower level (proven later).
These two differences cancel to result in the same stress patterns
for the $\NVvis$ and $\NVir$ transitions.
Conveniently this allows the visible and infrared spectra to be easily compared
to obtain the relative magnitudes of the $\NVvis$  and  $\NVir$ stress
parameters.
This is the intention of presenting Figure 2 where  spectra of $\NVvis$  and $\NVir$ are depicted for the same stress applied along the $\langle001\rangle$ and
$\langle111\rangle$ directions.

\begin{figure}
\begin{flushright}
\includegraphics[width=\textwidth]{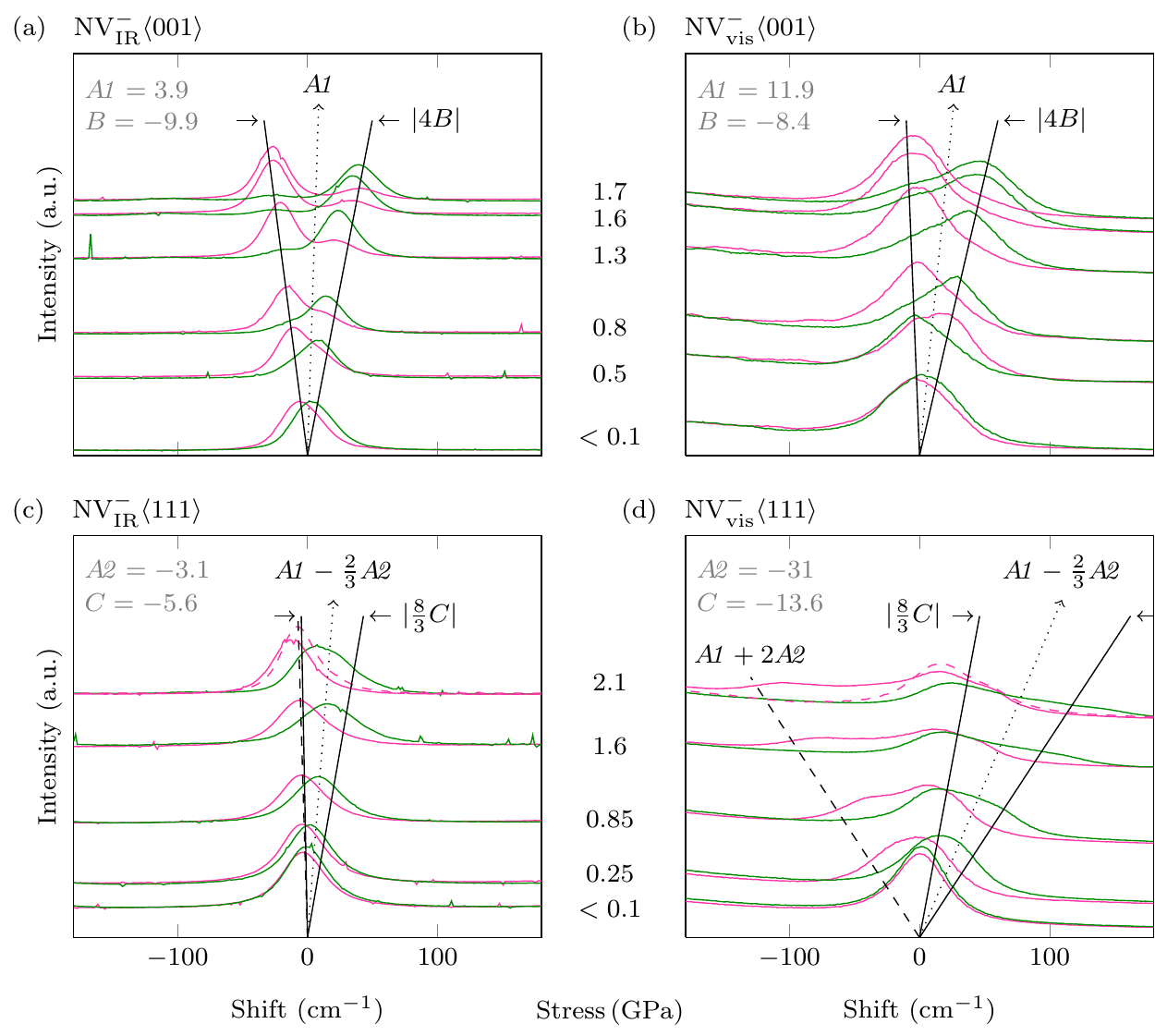}
\par\end{flushright}
\caption{
	Uniaxial stress spectra for $\NVir$ on the left and $\NVvis$ on the right.
	The upper traces (a) and (b) show spectra for $\langle001\rangle$ stress
	and the lower traces (c) and (d) show spectra for $\langle111\rangle$
	stress.
	Excitation was from 100\,mW laser at 532 nm, and emission was observed at
	right angles and recorded separately in $\pi$ (green) and $\sigma$
	(magenta) polarisations.
	Excitation polarisation was perpendicular to the stress direction
	($\sigma$) with the exception of the dashed traces in (c) and (d) where the
	laser polarisation was parallel to stress ($\pi$) and so the axial centres
	were not excited.
	The sample temperature was $\sim 150$\,K.
	Variation of stress across the sample prevented the lines from being well
	resolved (and breakage prevented improvement of the data).
	However, identical stress settings allow the relative size of the shifts
	and splittings to be compared between $\NVir$ and $\NVvis$.
	The straight lines and annotations indicate the stress parameters
	calculated later (not direct fits).
	The $\NVvis$ spectra are consistent with \cite{davies1976optical} and the
	stress parameters are those from reference \cite{davies1976optical}.
	}
 \end{figure}

For stress along $\langle001\rangle$ the splittings are the same for all
orientations of the $\mbox{NV}^{-}$ centre.
The ZPL is split into two components with one component $\sigma$ polarised
(electric field vector perpendicular to stress) and the other predominantly $\pi$
polarised (parallel to stress).
The splittings are determined by the value of the $B$ parameter (see Table 1)
and the average shift is given by $A\!\mathit{1}$.
It can be seen from comparing Figure 2(a) and (b) that $B\IR$ is marginally larger
than $B\vis$ whereas $A\!\mathit{1}\IR$ is only about one third of
$A\!\mathit{1}\vis$.

For $\langle111\rangle$ stress there are two subsets of centres (Table 1).
One subset contains the centres oriented along the stress direction, for which
there is no change of symmetry.
This means there is no splitting, but the transition is shifted by
$A\!\mathit{1} + A\!\mathit{2}$ (Table 1).
NV$^-$ centres in this orientation are not excited when the electric field vector
of the excitation is parallel to their axis, since the
$\A\!\leftrightarrow\!\E$ transitions do not involve a z dipole moment.
Consequently this orientation does not contribute to the dashed traces of
Figure 2 (c) and (d) where this excitation polarisation is adopted.
The NV$^-$ centres in this orientation do give a line when transverse excitation is
used.
This 'extra' line is barely discernible in the case of the infrared spectrum as
it overlaps the other features indicating a very small shift
($A\!\mathit{1}\IR$ + $A\!\mathit{2}\IR$).
In contrast, there is a large shift of this line for the visible transition.
Since $A\!\mathit{1}\vis$ and $A\!\mathit{1}\IR$ are known from the above
$\langle001\rangle$ stress measurements, it can be readily deduced that
$A\!\mathit{2}\vis$ for the visible is large and negative whereas
$A\!\mathit{2}\IR$ for the infrared is small.
This information is consistent with average shifts for the centres oriented at
70$^{\circ}$ to the $\langle111\rangle$ stress given by $A\!\mathit{1}+
\frac{2}{3} A\!\mathit{2}$ (Table 2).
The ZPL splitting for these centres depends on the $C$ parameter, and it is
apparent that $C\IR$ is about one third of $C\vis$.

The conclusion that $A\!\mathit{1}\IR$ and $C\IR$ are a factor of three smaller
than their $\NVvis$ counterparts is consistent with the $\NVir$ strain
parameters reported previously \cite{rogers2008infrared}.
However, there is no consistency with the $A\!\mathit{2}\IR$ and $B\IR$
parameters.
Here we have established that $A\!\mathit{2}\IR$ is an order of magnitude
smaller than $A\!\mathit{2}\vis$ (instead of the factor of 2.7 given
previously), and that $|B\IR| > |B\vis|$ (instead of the reverse).
The previous values relied on the interpretation of spectra for stress along
the $\langle110\rangle$ direction and, therefore, the spectra for this stress
direction are re-investigated in the next section.

\subsection{Uniaxial stress along $\langle110\rangle$ stress using selective excitation}

Stress along $\langle110\rangle$ causes the NV$^-$ centres to form two distinct
sets of orientations, both of which have some component of transverse strain
and therefore exhibit splitting (Table 1).
This produces a four-line structure in the spectrum, and the determination of
strain parameters depends heavily on a correct assignment of each line to a
transition in a given NV orientation.
Here we use selective excitation techniques to provide reliable assignments.

\begin{figure}
\hspace{2.3cm}
\includegraphics[width=109.8mm]{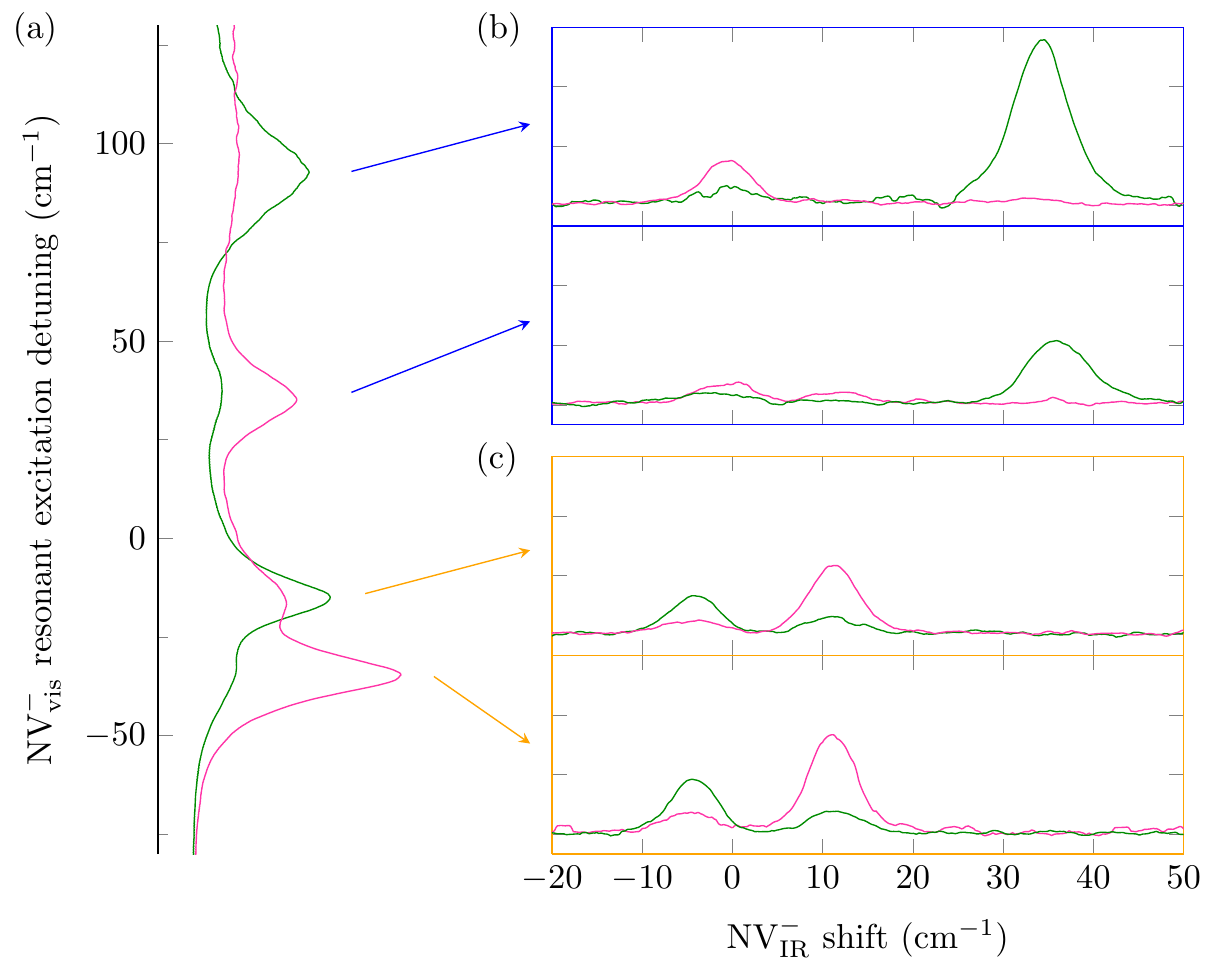}
\caption{
	Uniaxial $\langle110\rangle$ stress spectra using resonant $\NVvis$
	excitation to assign $\NVir$ peaks.
	The exciting laser and IR detection were along $\langle011\rangle$, and the
	sample temperature was 10\,K.
	(a) Excitation of the $\NVvis$ transition was obtained by sweeping the
	excitation laser between 640\,nm and 630\,nm and detecting emission within the
	vibrational band at $700\,\mbox{nm}$.
	(b)  IR spectra obtained with the excitation laser at fixed frequency
	resonant with the visible peak corresponding to the $\mbox{NV}^{-}$
	orientations \circled{3} and \circled{4} perpendicular to the stress (see
	Table 1).
	(c) The lower two traces framed in orange correspond to resonant excitation
	of orientations \circled{1} and \circled{2} at an angle of $36^{\circ}$ to
	the stress.
}
\end{figure}

A 200\,mW tunable dye laser was swept through the $\NVvis$ ZPL and the emission
was detected in the vibronic sideband between 650 nm and 750 nm.
Weak (1\,mW) $532$\,nm laser light was applied simultaneously to inhibit loss
of signal through hole burning.
The polarised $\NVvis$ excitation spectra for a $\langle110\rangle$ stress of
$1.4\,\mathrm{GPa}$ obtained in this way is shown in Figure 3(a).
This $\NVvis$ excitation spectrum is consistent with the measurements of Davies
and Hamer obtained in absorption \cite{davies1976optical}.
The two higher energy lines in excitation (at 632.4\,nm and 634.8\,nm) are
associated with centres at right angles to the stress (orientations \circled{3}
and \circled{4}, Table 1), and the lower energy $\NVvis$ lines are associated
with orientations \circled{1} and \circled{2} which are at $36^\circ$ to the
stress \cite{davies1976optical}.

The detection filter was changed to measure emission in the $\NVir$ band.
Resonantly exciting the two higher energy $\NVvis$ transitions gave the
polarised $\NVir$ spectra shown in Figure 3(b).
Since these laser frequencies only excite the orientations \circled{3} and
\circled{4} which are orthogonal to the stress, the $\NVir$ spectrum shows only
two lines.
These lines are clearly either predominantly $\pi$ or $\sigma$ polarised,
enabling them to be assigned to the $\Gamma_1$ and $\Gamma_2$ components
according to Table 1.
Tuning the laser to the lower energy $\NVvis$ transitions caused only
orientations \circled{1} and \circled{2} to be excited, producing the $\NVir$
spectra shown in Figure 3(c).
Again the lines are strongly polarised and readily assigned using Table 1.
There is always the equivalence between the visible and infrared spectra
described in previous section but it is noted that the order of the $\pi$ and
$\sigma$ lines for the 36$^{\circ}$ case are reversed between the visible and
infrared spectra.
This results from a reversal of the relative strengths of the $B$ and $C$
stress parameters between the visible and infrared cases with $B\IR > C\IR$ in
one case and $C\vis > B\vis$ in the other.

These selective excitation measurements provide the first unambiguous
assignments of the infrared spectral features for $\langle110\rangle$ stress.
It is now clear that the significantly different $A\!\mathit{2}\IR$ and $B\IR$
spitting parameters given previously \cite{rogers2008infrared} resulted from an
incorrect assignment of the $\NVir$ lines for $\langle110\rangle$ stress.
In that work it was assumed that the four peaks were in the same order as for
$\NVvis$, which does not turn out to be the case.

Having established the identity of each line in the spectrum, more conventional
photoluminescence (PL) measurements were made using the 532\,nm non-resonant
excitation.
In this way the position of the four lines in the stress spectra were followed
for stress values in the range 0--3\,GPa, and the shifts and splittings are shown in
Figure 4.
This figure also includes the results for stress along $\langle001\rangle$ and
$\langle111\rangle$, where there is less ambiguity in the assignments of
the lines and therefore no advantage to adopting selective excitation
techniques.
It can be seen from the figures that the displacements with stress are not
always linear and this requires consideration before the values of the stress
parameters can be deduced.

\begin{figure}
\begin{flushright}
\includegraphics[width=159.4mm]{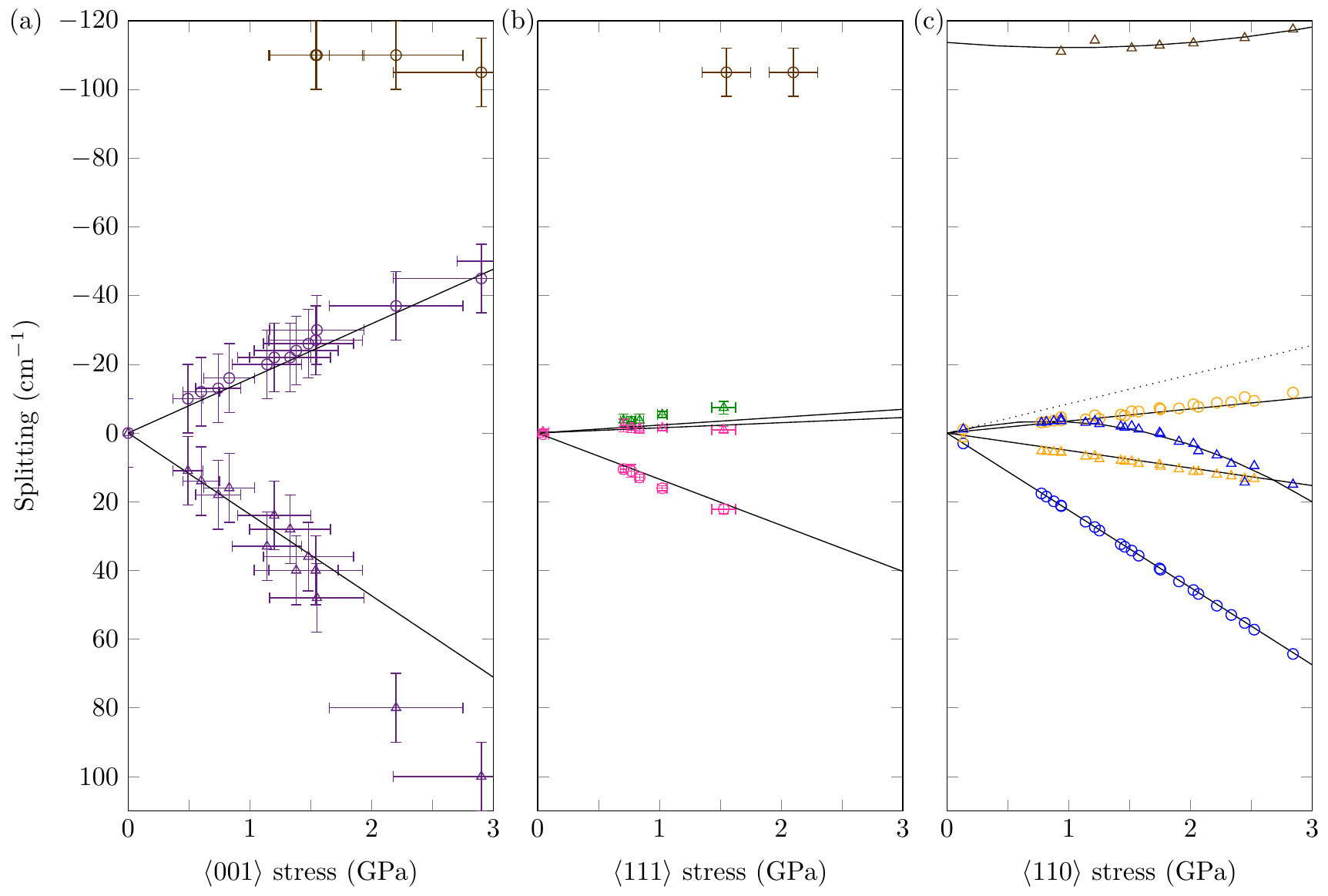}
\end{flushright}
\caption{
	Strain splitting of the IR line for
	(a) $\langle110\rangle$,
	(b)
	$\langle001\rangle$ and
	(c) $\langle111\rangle$.
	The vertical scale is reversed to correspond to the emission spectrum where
	the lower level splits.
	The spectra were measured independently in $\pi$ (circles) and $\sigma$
	(triangles) polarisation.
	For each stress direction the data points are coloured to match the sets of
	NV orientations given in Table 1.
	The error in stress is large in (a) due to the sample breaking over the
	course of the measurement.
	}
\end{figure}

\subsection{High stress and extra feature at 115\,cm$^{-1}$}

\begin{figure}
\centering{\includegraphics[width=109.8mm]{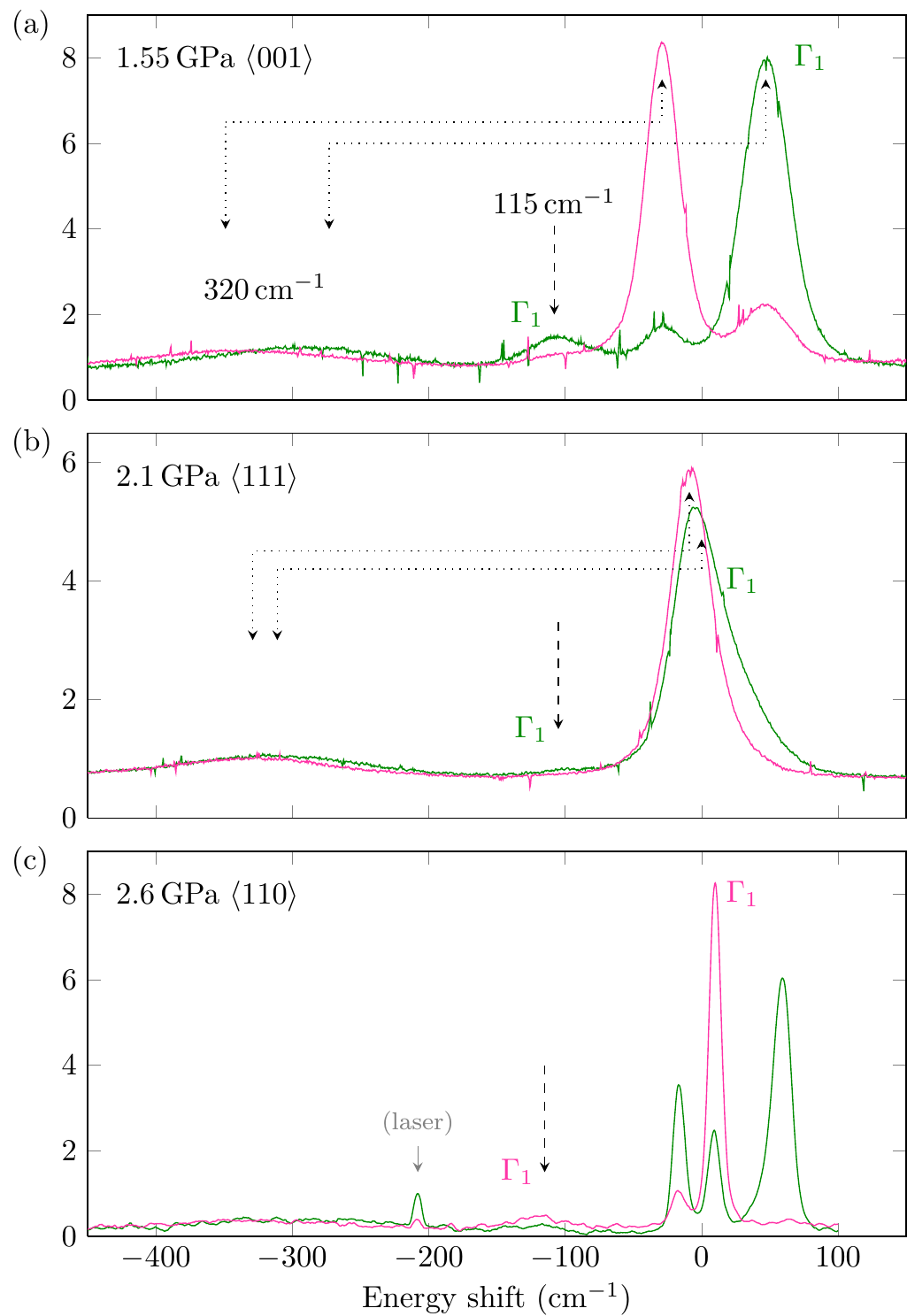}}
\caption{
	Uniaxial stress spectra including vibrational features.
	The upper trace (a)
	gives the spectra for $\langle001\rangle$ stress and the central trace (b) for
	$\langle111\rangle$ stress. In these cases the sample temperature was 150\,K.
	(c) For $\langle110\rangle$ stress the sample
	temperature was 10\,K and the higher resolution was obtained by detecting
	emission from a small volume using masking. In all cases the feature at 115\,cm$^{-1}$ is induced by the stress. It has the same polarisation as the
	$\Gamma_1$ component of the ZPL ( $\pi$ in the upper two traces and
	$\sigma$ in lowest trace - see Table 1). The first vibrational sideband at
	320 cm$^{-1}$ can be seen to have the same polarisation as the ZPL
	indicating the vibration has A$_1$ symmetry.
	}
\end{figure}

At higher stress ($>1$\,GPa) an extra feature was found to be induced  $115
\pm5\,\mbox{cm}{}^{-1}$ to the low energy side of the ZPL and this is shown in
Figure 5 for $\langle111\rangle$, $\langle110\rangle$, and $\langle001\rangle$
stress.
The feature appears with varying intensities but increases in strength with
stress at the expense of a component of the ZPL with the same polarisation.
The measurements were made at higher resolution for the case of
$\langle110\rangle$ uniaxial stress, and it is apparent that the extra feature
gains at the expense of the line displaced non-linearly.
It also shifts slightly in the reverse direction, as shown in Figure 4.
This is typical for a situation where there are two interacting levels which
have the same symmetry.
From the analysis of the ZPL it has been established that the line shifting
non-linearly has $\Gamma_1$ in $\Cs$ symmetry.
The extra feature will, therefore, also have $\Gamma_1$ in $\Cs$ and since it
is not split it must have $\Aone$ symmetry in $\Cthreev$.

This symmetry assignment is consistent with its occurrence for other stress
directions.
For $\langle001\rangle$ stress the line displaced to higher energy has $\pi$
polarisation and is assigned to a $\Gamma_1$ state, and this line mixes with
the extra feature (Figure 5(a)).
Even though the interacting ZPL component is shifting away from the 115
cm$^{-1}$ feature, the displacement of this ZPL line becomes non-linear as a
result of the interaction, as shown in Figure 4(a).
Here it might be expected that the 115 cm$^{-1}$ feature shifts in the reverse
direction, but the effect is reduced owing to the proximity of the
320\,cm$^{-1}$ vibrational level.
Indeed at the highest stress it is observed that there is a slight shift of the
extra feature to shorter wavelength (higher energy) owing to the latter
interaction.
In the third case of $\langle111\rangle$ the effects are small but the feature
again has the same polarisation as that for the $\Gamma_1$ component of the
split ZPL (Figure 5(b)).

Since the 115 cm$^{-1}$ feature interacts with one component of a line that
splits with stress, it must be associated with the $\singE$ electronic state.
It occurs on the low energy side of the ZPL in the emission spectrum.
Should the $\singE$ be the upper singlet level there will be relaxation to this
level 115 cm$^{-1}$ below the $\singE$ state and at cryogenic temperatures
($<30$\,K) all the emission would be from this level.
This is not the case and it is concluded that the $\singE$ is not the upper
singlet level.
The alternative is that the $\singE$ is the lower singlet level and the extra
level lies 115\,cm$^{-1}$ above it.
This confirms our previous report
\cite{manson2010optically} and is consistent with the now generally accepted
theoretical model \cite{doherty2011negatively, maze2011properties}.

The occurrence of low-energy vibronic levels in diamond is a fairly common observation
and has been observed in previous uniaxial stress studies of diamond
\cite{davies1979dynamic}.
They are associated with a dynamic Jahn-Teller effect associated with an $\E$ state.
Davies \cite{davies1979dynamic} has established five other cases of trigonal
centres in diamond exhibiting this effect.
The first vibrational state associated with a degenerate $\E$ vibration will
involve the electronic and the vibration states, resulting in four vibronic
states with symmetries $\E \times \E = \Aone + \Atwo + \E$.
The $\E$ vibronic level is displaced up in energy and the $\Aone + \Atwo$ down.
Quadratic electron-vibration interaction will lift the $\Aone$ and $\Atwo$
degeneracy and result in the low lying $\Aone$ state at 115\,cm$^{-1}$ as
observed here.
A similar situation arises in the case of the $\doubE$ ground state of
$\mbox{NV}^{0}$ \cite{davies1979dynamic}.
In this case the level occurs at 110 cm$^{-1}$ and from the similarity in the
situation it can be expected that the strength of the Jahn-Teller interaction
is similar: E$_{JT}$/$\hbar\omega \sim 2$.

This vibronic level has significant implications.
It has been used above to establish the order of the singlet levels, resolving
long-standing contention about this detail of the NV electronic structure
\cite{goss1996twelve-line, manson2007issues, gali2008ab, rogers2008infrared,
delaney2010spin-polarization}.
It should also be noted that one consequence of the dynamic Jahn-Teller
interaction is a reduction of the effect of perturbations
\cite{ham1972jahn-teller}.
Thus the experimental measurements of the stress splittings will be slightly
smaller than obtained from calculation unless such effects are included.

\subsection{Stress parameters}

The non-linear shift of some spectral features with stress is, therefore, due
to interaction with vibronic levels.
Modelling these interactions is not straightforward as they can involve a
distribution of vibrations and the distribution need not be simple.
Consequently we have determined the stress parameters using shifts and
splitting at levels of stress where the strength of this latter interaction is
negligible, essentially using the asymptotic slopes at zero stress.
The value of the parameters are given in Table 2.

\begin{table}
\caption{Stress parameters of the $\NVir$ ZPL compared to those
for $\NVvis$ , $\mbox{NV}^{0}$ and N3, given in $\mbox{cm}^{-1}/\mbox{GPa}$ (meV/$\mbox{GPa}$) all $\A\! \leftrightarrow\! \E$ transitions at trigonal vacancy centres with adjacent nitrogen atoms. The values
for $\NVvis$ are taken from \cite{davies1976optical} although $B$ and $C$ sign change appropriate for $\Atwo\! \leftrightarrow\! \E$ transition. The values for NV$^0$ are from \cite{davies1979dynamic} and N3 from \cite{crowther1967phonon}. }
\label{tab:stressparameters}. \hspace{24mm}
\onehalfspacing
\small
\centering{}%
\begin{tabular}{ccc|c|cc}
\hline
 & $\NVir$  & $\NVvis$   & $\NVir/\NVvis$  & $\mbox{NV}^0$ & N3 \tabularnewline
 &$1042.6\,\mbox{nm}$ & $637\,\mbox{nm}$ &ratio & 575\,\mbox{nm} & 415\,\mbox{nm}\tabularnewline
\hline
Param (pert) & cm$^{-1}$(meV) & cm$^{-1}$ (meV) &       &cm$^{-1}$(meV) & cm$^{-1}$(meV)\tabularnewline\hline
$A\!\mathit{1}\ \left( \Aone \right)$  & 3.9 $\pm0.3$ (0.48) & 11.9 (1.47)  & 0.33 & 8.5  (1.05) & 4.0 (0.5)\tabularnewline
$A\!\mathit{2}\ \left( \Aone^{'} \right)$  & -3.1$\pm0.3$ (-0.38)   & -31.0 (-3.85)  & 0.10  & -28.6 (-3.55) & 34 (4.2)  \tabularnewline
$B\ \left( \E \right)$  & -9.9 $\pm0.5$ (-1.23)   & -8.38 (-1.04)  & 1.2 & 12.5 (1.55) & -8.5 (-1.55)  \tabularnewline
$C\ \left( \E^{'} \right)$  & -5.6 $\pm0.5$ (-0.69)  & -13.6 (-1.69)  & 0.41 & 14.1 (1.76) & -11 (-1.9)  \tabularnewline
\hline
\end{tabular}
\end{table}

\section{Discussion of the molecular model}

\begin{figure}
\begin{centering}
\includegraphics[width=1.0\columnwidth]{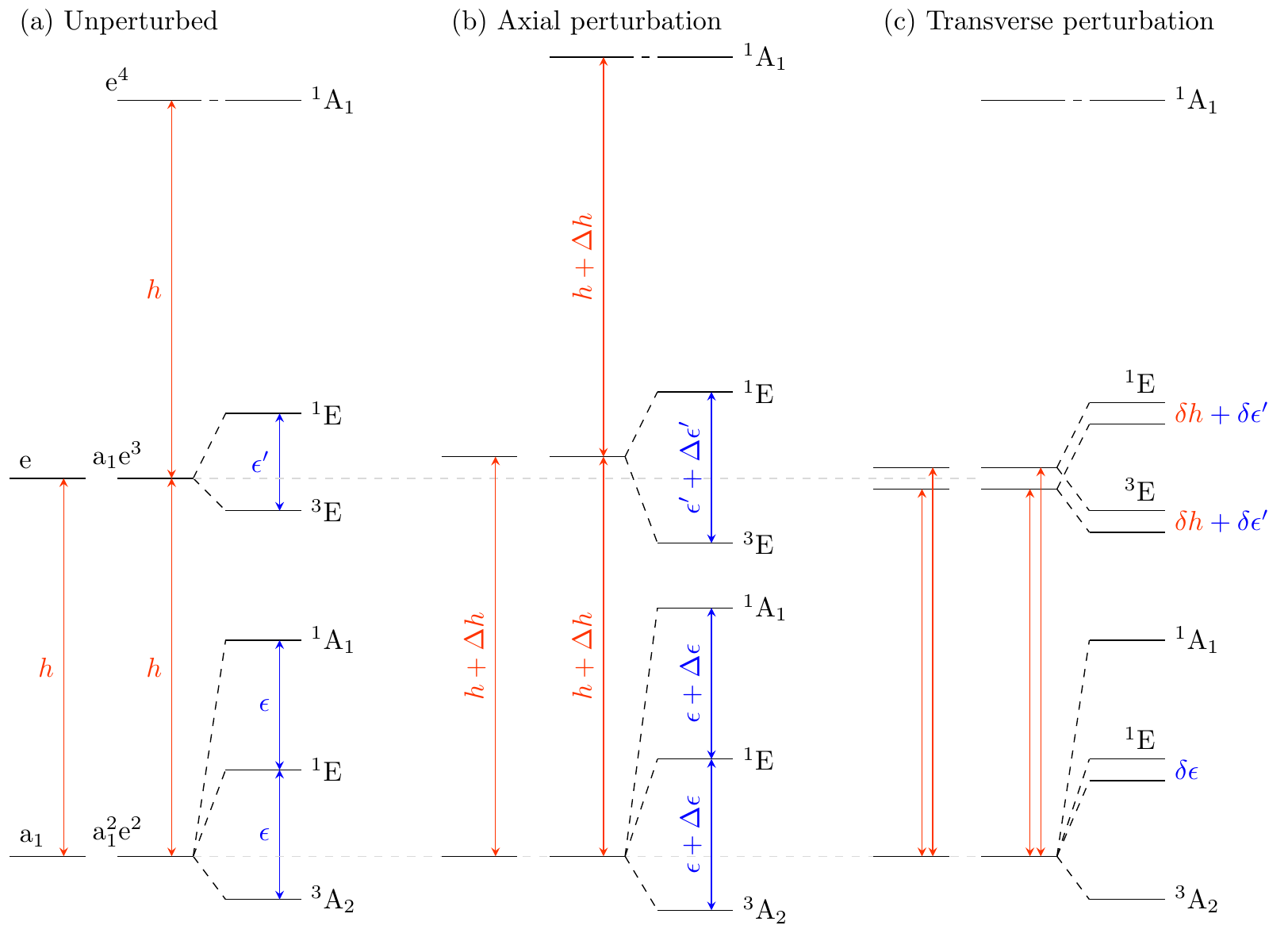}
\par\end{centering}
\caption{
	Electronic energy level scheme described by the molecular model for three
	situations. 
	For each situation, the molecular orbital energy levels are depicted on the
	left, the configuration energy levels in the centre and the multi-electron
	state energy levels on the right. 
	Effects arising from one-electron Coulomb interaction are coloured red, and
	effects arising from electron-electron interaction are blue.
	(a) In the unperturbed case the molecular orbitals $\ai$ and $\e$ are
	separated by energy $h$ due to the one-electron Coulomb interaction.
	The three configurations associated with four electrons occupying these
	molecular orbitals are therefore also separated by $h$.
	The singlet and triplet energy levels within each configuration are
	separated by the electron-electron Coulomb repulsion interaction.
	In first order the three levels of the $\ai^2\e^2$ configuration are
	equally separated by $\epsilon$.
	(b) Axial strain produces a perturbed trigonal symmetry, which results in
	changes $\Delta h$ and $\Delta\epsilon$ to the separations $h$ and
	$\epsilon$. Since trigonal symmetry is maintained, this distortion does not
	split the E states.
	(c) Transverse strain lowers the symmetry and gives rise to the splittings
	$\delta h$, $\delta\epsilon$ and $\delta\epsilon^\prime$.
	Note in this case the only configuration that is split by the one-electron
    Coulomb	interaction is $\ai\e^3$ because it has an odd numbers of electrons occupying the $\e$ molecular orbitals.
	The electron-electron Coulomb repulsion interaction can split all degenerate
	levels.
	The two $\singE$ states and the two $\singAone$ states can mix through
	electron-electron Coulomb repulsion.

	}
\end{figure}

The electronic model of the $\mathrm{NV}^-$ centre has its foundations in the
defect-molecule approach of Coulson and Kearsley \cite{coulson1957colour} and
has been given in detail by many authors
\cite{lenef1996electronic,gali2008ab,doherty2011negatively,
maze2011properties,doherty2013nitrogen-vacancy}.
The centre's electronic states are written in terms of symmetry-adapted
molecular orbitals.
There are four unbound sp$^3$ atomic orbitals adjacent to the vacancy and in
$\Cthreev$ symmetry these can be linearly combined to give two degenerate
orbitals that transform as the $\E$ irreducible representation (denoted as
$\e$-orbitals) and two separate orbitals of A$_1$ symmetry (denoted as
$\ai$-orbitals).
These are occupied by six electrons: one from each of the adjacent carbon
atoms, two from the nitrogen, and one acquired from the lattice.
The lower $\ai$ orbital is always occupied and need not be included in a
description of the states.
The occupancy of the other four electrons describe the multi-electron states.

The non-relativistic electronic Hamiltonian of the $\mathrm{NV}^-$ centre may
be defined as \cite{doherty2011negatively} \begin{eqnarray} H(\vec{r},\vec{R})
= T_e(\vec{r})+V_\mathrm{Ne}(\vec{r},\vec{R})+V_\mathrm{ee}(\vec{r}),
\end{eqnarray} where $T_e$ is the electronic kinetic energy, $V_\mathrm{Ne}$ is
the one-electron Coulomb interaction between the $\NV$ electrons and the
lattice nuclei and electrons, $V_\mathrm{ee}$ is the electron-electron Coulomb
repulsion interaction of the $\NV$ electrons, $\vec{r}$ are the collective
coordinates of the $\NV$ electrons and $\vec{R}$ are the collective coordinates
of the lattice.
Both $T_e$ and $V_\mathrm{Ne}$ can be written as sums of one-electron
operators, whereas $V_\mathrm{ee}$ can be written as a sum of two-electron
operators.
The molecular orbitals are defined as solutions of the one-electron terms
$T_e+V_\mathrm{Ne}$.
The $\ai$ and $\e$ molecular orbitals have energies that lie within the diamond
band gap and are separated by $h\sim$ 2\,eV (Figure 6(a)).
The energies of the $\ai$ and $\e$ molecular orbitals define the energies of
the electronic configurations.
The four electrons occupying these molecular orbitals lead to three
configurations $\ai^{2}\e^{2}$, $\ai \e^{3}$ and $\e^{4}$, which are each
separated by $h$ (Figure 6(a)).
The introduction of the electron-electron Coulomb interaction $V_\mathrm{ee}$
separates the multi-electron states within a configuration into triplet and
singlet levels.
The separation can be of the order of eV and, hence, comparable in magnitude to
that of the one-electron terms.
For example, the lowest energy configuration $\ai^{2}\e^{2}$ is split into
equally separated states $\tripAtwo$, $\singE$ and $\singAone$ with separations
of $\epsilon\sim$ 1\,eV \cite{doherty2011negatively, maze2011properties}
(Figure 6(a)).
The electron-electron Coulomb repulsion interaction can also give interaction
between configurations and mix the singlet levels of the same orbital symmetry,
thereby modifying the simple expressions for $\epsilon$ and $\epsilon^\prime$.

The above one- and two-electron Coulomb interactions give the dominant terms in
determining the effects of stress, which are observed to be several meV.
Other electronic interactions such a spin-orbit and spin-spin are less than meV
and their effects are negligible compared to stress.
Whilst electron-vibration interaction can be of the order of meV, it can not
give stress splitting by itself.
Although, as mentioned earlier, it can modify the magnitude of stress splittings in the case
of dynamic Jahn-Teller interaction \cite{ham1972jahn-teller}.
Hence, the analysis of the stress parameters can be largely restricted to
consideration of the Coulomb interactions.

When stress is applied, the lattice coordinates $\vec{R}$ change, which results
in a change $\delta V_\mathrm{Ne}[s]$ of the one-electron Coulomb interaction
that in turn modifies the molecular orbitals and their energies.
Furthermore, the modification of the molecular orbitals occupied by the
electrons leads to a change in the electron-electron Coulomb repulsion
interaction, which can be represented by the effective operator $\delta
V_\mathrm{ee}[s]$.
Note that $\delta V_\mathrm{Ne}[s]+\delta V_\mathrm{ee}[s]$ can be expanded in
symmetry adapted form with terms that are in one to one correlation with those
in equation (1).
If the symmetry is not changed by the applied stress, $\delta V_\mathrm{Ne}[s]$
will only alter the $\ai$--$\e$ energy separation and this is denoted by
$\Delta h$ (Figure 6(b)).
This will result in a change of the energy separation between configurations
but cause no change within each configuration (Figure 6(b)).
Where the applied stress lowers the symmetry of the centre, $\delta
V_\mathrm{Ne}[s]$ will result in a splitting of the $\e$ molecular orbitals by
$\delta h$.
The consequence is that the multi-electron $\E$ states with an odd number of
electrons occupying the $\e$ molecular orbitals will be split by $\delta h$
(Figure 6(c)).
Significantly, no splitting occurs when an even number of electrons occupy the
$\e$ molecular orbitals.
For each pair of electrons, one of the $\e$ electrons is moved up in energy and
the other down, such that there is no overall splitting.

It is convenient to first consider the $\delta V_\mathrm{Ne}[s]$ interaction in
relation to the singlet transition.
The $\singAone (\ai^2 \e^2) \leftrightarrow \singE (\ai^2 \e^2)$ transition is
between levels within the same $\ai^2 \e^2$ configuration and so the transition
energy can not be shifted by $\delta V_\mathrm{Ne}[s]$.
In addition, the $\singE(\ai^2 \e^2)$ state has an even number of electrons
occupying the $\e$ molecular orbitals and so there will be likewise no
splitting arising from $\delta V_\mathrm{Ne}[s]$.
Mixing between singlet levels can change this situation.
However, if this mixing was predominately responsible for the observed stress
response of the singlet transition, the ratio of the $A\!\mathit{1}$ parameter
for $\NVvis$ and $\NVir$ would be same as for the $A\!\mathit{2}$ parameter.
The $A\!\mathit{1}$ ratio is 0.33 and that of $A\!\mathit{2}$ is 0.1 (Table 2)
and, hence, the experimental shifts of the singlet transition can not be simply
explained by the $\delta V_\mathrm{Ne}[s]$ interaction, even when allowing for
mixing of the singlets.
Likewise, the $\delta V_\mathrm{Ne}[s]$ interaction with mixing would only be
able to account for small splitting of $\singE(\ai^2 \e^2)$ compared to that
for $\tripE(\ai \e^3)$, whereas the splitting parameters $B$ and $C$ for
	$\NVvis$ and $\NVir$ are of comparable size (Table 2).
The dominant interaction giving rise to the stress shift and splitting of the
singlet transition at 1042\,nm  must result from an alternative interaction.
The most obvious candidate, given the magnitude of this interaction, is
electron-electron Coulomb repulsion interaction $\delta V_\mathrm{ee}[s]$.
The first order changes are taken to be $\Delta\epsilon$ and $\delta\epsilon$
for axial and transverse stress, respectively (Figure 6(b) and (c)).
Such perturbations can account for the change of the $\singAone(\ai^2 \e^2)$ to
$\singE(\ai^2 \e^2)$ separation and the splitting of the $\singE(\ai^2 \e^2)$
level.

The situation for the  $\tripAtwo(\ai^2 \e^2) \leftrightarrow \tripE(\ai \e^3)$
triplet transition is very different.
The transition is between states of different configuration and the $\tripE(\ai
\e^3)$ state has an odd number of electrons occupying the $\e$ molecular
orbitals.
Consequently, the shifts and splitting of the $\NVvis$ can arise as a
consequence of the changes of the one-electron Coloumb interaction $\delta
V_\mathrm{Ne}[s]$.
However, the possibility that there are contributions from $\delta
V_\mathrm{ee}[s]$ cannot immediately be eliminated.
To determine how much this latter term contributes, it is worth considering the
situation for $\mbox{NV}^{0}$.

The $\mbox{NV}^{0}$ centre has one less electron and its transition is between
a $\doubE(\ai^2 \e)$ ground state and a ${}^2\mathrm{A}_2 (\ai \e^2)$ excited
state \cite{manson2013assignment}.
This $\doubE$ ground state has an odd number of electrons occupying the $\e$
molecular orbitals and hence can be split as a result of the $\delta
V_\mathrm{Ne}[s]$ interaction.
It is also the sole state of the $\ai^2 \e$ configuration and so there can be
no contribution from $\delta V_\mathrm{ee}[s]$.
Therefore, the splitting of the $\mbox{NV}^{0}$ ZPL at 575 nm must arise solely
from the $\delta V_\mathrm{Ne}[s]$ interaction.
The splitting is that of a single $\e$-electron and is expected to be of
similar magnitude (but opposite sign) to that of the single $\e$-hole in the
case of $\tripE(\ai \e^3)$.
From Table 2 it is clear that the $B$ and $C$ parameters have similar magnitude
for $\NVvis$ and $\mbox{NV}^{0}$, which is consistent with this expectation.
This provides strong evidence that the dominant contribution to the splitting
of the $\NVvis$  transition arises from the $\delta V_\mathrm{Ne}[s]$
interaction, and any contribution from $\delta V_\mathrm{ee}[s]$ is minor.

The $\mbox{NV}^{-}$ centre is the first colour centre in diamond where the
stress parameters are known for two separate transitions, and this provides an
ideal situation for testing theoretical calculations.
For example it may help determine whether the separate contributions from
one-electron Coulomb interaction and electron-electron Coulomb repulsion
interaction, as outlined above, can be justified.
Having similar information for a transition in the closely related neutral
charge state $\mbox{NV}^{0}$ \cite{davies1979dynamic} is also valuable.
The $\mbox{N3}$ centre is another nitrogen-related colour centre that has been
studied by uniaxial stress \cite{crowther1967phonon}.
It involves three nitrogen atoms and one carbon adjacent to a vacancy, rather
than the three carbon and one nitrogen, and a similar molecular model is
adopted for describing its electronic states.
Despite the stress parameters being similar to those of $\NVvis$, in this case
the molecular model has not successfully predicted all of the excited states
\cite{jones1997n2}.
Having the stress parameters for four related transitions as given in Table 2
provides valuable information for {\em ab initio} calculations to test our
understanding of the electronic model of nitrogen-related colour centres in
diamond.

\section{Summary and conclusions}

The aim of the work was to use uniaxial stress techniques to better understand
the singlet levels of the nitrogen-vacancy centre in diamond.
The 1042\,nm zero-phonon line is understood to be associated with the singlet
to singlet transition between levels in the same configuration.
The ZPL is spectrally narrow, the sideband is weak, and the symmetry
maintaining stress shift parameters $A\!\mathit{1}$ and $A\!\mathit{2}$ are also
relatively small and these are all characteristics of a transition between
levels in the same electronic configuration.
The $\singAone(\ai^2 \e^2) \leftrightarrow \singE(\ai^2 \e^2)$ singlet-singlet
transition is the only transition within the electronic model that satisfies
this condition and these aspects all give confidence that the transition is
correctly identified.
However, the stress splitting parameters are large and
comparable with those for the $\A\! \leftrightarrow\! \E$ triplet and doublet transitions
of $\NVvis$ and $\mbox{NV}^{0}$, respectively.
These latter transitions involve a change of configuration and an $\E$ state
with an odd number of $\e$ electrons. Consequently, one-electron Coulomb interaction can
account for such effects.
The singlet-singlet transition is different since the one-electron Coulomb
interaction can not (in first order) split or shift the ZPL, and so it was
anticipated the responses would be smaller.
The strain parameters for $\NVir$ must arise from an alternative interaction
and in this work it has been shown that they can be attributed to the
two-electron Coulomb repulsion term.
It is recognised within the Coulson and Kearsley \cite{coulson1957colour} model
that Coulomb repulsion always plays a significant role and in the case of the
NV$^-$ centre this interaction is of comparable magnitude to the one-electron
Coulomb term.
It is, therefore, realistic that the one-electron and two-electron Coulomb
interactions can result in similar energy changes in response to a distortion
of the lattice.
The conclusion is that there is overall consistency with the current electronic
model of the NV$^-$ and it follows that there is an adequate understanding of
the singlet states.

The present uniaxial stress studies have also established that there is a
dynamic Jahn-Teller effect associated with the $\singE$ level.
Combining this observation with previous reports of dynamic Jahn-Teller effect
in the excited $\tripE$ state, it is clear that electron-vibration interaction
is significant within the NV$^{-}$ system.
The presence of electron-vibration interaction has been determined from
observations within the $\singE$  and $\tripE$ degenerate electronic states
independently, but the interaction can have more significant consequences
between states.
In particular, it can play a role in inter-system crossing between  $\tripE$ and
$\singAone$ and between $\singE$ and $\tripAtwo$ triplet and play a very
important role in giving rise to the important spin polarisation property of
NV${}^-$.

\section*{Acknowledgements}
\addcontentsline{toc}{section}{Acknowledgements}
This work was supported by the Australian Research Council (DP 120102232).

\section*{References}

\bibliographystyle{unsrt}
\addcontentsline{toc}{section}{\refname}\bibliography{nv_singlet_levels}

\begin{thebibliography}{10}

\bibitem{doherty2013nitrogen-vacancy}
Marcus~W. Doherty, Neil~B. Manson, Paul Delaney, Fedor Jelezko, J{\"o}rg
  Wrachtrup, and Lloyd~C.L. Hollenberg.
\newblock The nitrogen-vacancy colour centre in diamond.
\newblock {\em Physics Reports}, 528(1):1 -- 45, 2013.

\bibitem{chernobrod2005spin}
Boris~M. Chernobrod and Gennady~P. Berman.
\newblock Spin microscope based on optically detected magnetic resonance.
\newblock {\em Journal of Applied Physics}, 97(1):014903, 2005.

\bibitem{degen2008scanning}
C.~L. Degen.
\newblock Scanning magnetic field microscope with a diamond single-spin sensor.
\newblock {\em Appl. Phys. Lett.}, 92:243111, 2008.

\bibitem{taylor2008high-sensitivity}
J.~M. Taylor, P.~Cappellaro, L.~Childress, L.~Jiang, D.~Budker, P.~R. Hemmer,
  A.~Yacoby, R.~Walsworth, and M.~D. Lukin.
\newblock High-sensitivity diamond magnetometer with nanoscale resolution.
\newblock {\em Nature Physics}, 4:810 -- 816, 2008.

\bibitem{maze2008nanoscale}
J.~R. Maze, P.~L. Stanwix, J.~S. Hodges, S.~Hong, J.~M. Taylor, P.~Cappellaro,
  L.~Jiang, M.~V.~Gurudev Dutt, E.~Togan, A.~S. Zibrov, A.~Yacoby, R.~L.
  Walsworth, and M.~D. Lukin.
\newblock Nanoscale magnetic sensing with an individual electronic spin in
  diamond.
\newblock {\em Nature}, 455:644 -- 647, 2008.

\bibitem{balasubramanian2008nanoscale}
Gopalakrishnan Balasubramanian, I.~Y. Chan, Roman Kolesov, Mohannad Al-Hmoud,
  Julia Tisler, Chang Shin, Changdong Kim, Aleksander Wojcik, Philip~R. Hemmer,
  Anke Krueger, Tobias Hanke, Alfred Leitenstorfer, Rudolf Bratschitsch, Fedor
  Jelezko, and J{\"o}rg Wrachtrup.
\newblock Nanoscale imaging magnetometry with diamond spins under ambient
  conditions.
\newblock {\em Nature}, 455:648 -- 651, 2008.

\bibitem{hall2009sensing}
L.~T. Hall, J.~H. Cole, C.~D. Hill, and L.~C.~L. Hollenberg.
\newblock Sensing of fluctuating nanoscale magnetic fields using
  nitrogen-vacancy centers in diamond.
\newblock {\em Physical Review Letters}, 103(22):220802, November 2009.
\newblock Copyright (C) 2010 The American Physical Society; Please report any
  problems to prola@aps.org.

\bibitem{cole2009scanning}
Jared~H Cole and Lloyd C~L Hollenberg.
\newblock Scanning quantum decoherence microscopy.
\newblock {\em Nanotechnology}, 20(49):495401, December 2009.

\bibitem{hall2010ultra-sensitive}
Liam~T Hall, Charles~D Hill, Jared~H Cole, and Lloyd C.~L Hollenberg.
\newblock Ultra-sensitive diamond magnetometry using optimal dynamic
  decoupling.
\newblock {\em 1003.3699}, March 2010.

\bibitem{fu2007characterization}
Chi-Cheng Fu, Hsu-Yang Lee, Kowa Chen, Tsong-Shin Lim, Hsiao-Yun Wu, Po-Keng
  Lin, Pei-Kuen Wei, Pei-Hsi Tsao, Huan-Cheng Chang, and Wunshain Fann.
\newblock Characterization and application of single fluorescent nanodiamonds
  as cellular biomarkers.
\newblock {\em Proceedings of the National Academy of Sciences}, 104(3):727 --
  732, January 2007.

\bibitem{chang2008mass}
Yi-Ren Chang, Hsu-Yang Lee, Kowa Chen, Chun-Chieh Chang, Dung-Sheng Tsai,
  Chi-Cheng Fu, Tsong-Shin Lim, Yan-Kai Tzeng, Chia-Yi Fang, Chau-Chung Han,
  Huan-Cheng Chang, and Wunshain Fann.
\newblock Mass production and dynamic imaging of fluorescent nanodiamonds.
\newblock {\em Nat. Nanotechnol.}, 3:284--288, 2008.

\bibitem{tisler2009fluorescence}
Julia Tisler, Gopalakrishnan Balasubramanian, Boris Naydenov, Roman Kolesov,
  Bernhard Grotz, Rolf Reuter, Jean-Paul Boudou, Patrick~A. Curmi, Mohamed
  Sennour, Alain Thorel, Michael B{\"o}rsch, Kurt Aulenbacher, Rainer Erdmann,
  Philip~R. Hemmer, Fedor Jelezko, and J{\"o}rg Wrachtrup.
\newblock Fluorescence and spin properties of defects in single digit
  nanodiamonds.
\newblock {\em {ACS} Nano}, 3(7):1959--1965, July 2009.

\bibitem{gaebel2006room-temperature}
Torsten Gaebel, Michael Domhan, Iulian Popa, Christoffer Wittmann, Philipp
  Neumann, Fedor Jelezko, James~R. Rabeau, Nikolas Stavrias, Andrew~D.
  Greentree, Steven Prawer, Jan Meijer, Jason Twamley, Philip~R. Hemmer, and
  Jorg Wrachtrup.
\newblock Room-temperature coherent coupling of single spins in diamond.
\newblock {\em Nature Physics}, 2:408, June 2006.

\bibitem{dutt2007quantum}
M.~V.~Gurudev Dutt, L.~Childress, E.~Togan L.~Jiang, J.~Maze, F.~Jelezko, A.~S.
  Zibrov, P.~R. Hemmer, and M.~D. Lukin.
\newblock Quantum register based on individual electronic and nuclear spin
  qubits in diamond.
\newblock {\em Science}, 316:1312--1316, June 2007.

\bibitem{togan2010quantum}
E.~Togan, Y.~Chu, A.~S. Trifonov, L.~Jiang, J.~Maze, L.~Childress, M.~V.~G.
  Dutt, A.~S. Sorensen, P.~R. Hemmer, A.~S. Zibrov, and M.~D. Lukin.
\newblock Quantum entanglement between an optical photon and a solid-state spin
  qubit.
\newblock {\em Nature}, 466(7307):730--734, 2010.

\bibitem{neumann2010quantum}
P.~Neumann, R.~Kolesov, B.~Naydenov, J.~Beck, F.~Rempp, M.~Steiner, V.~Jacques,
  G.~Balasubramanian, M.~L. Markham, D.~J. Twitchen, S.~Pezzagna, J.~Meijer,
  J.~Twamley, F.~Jelezko, and J.~Wrachtrup.
\newblock Quantum register based on coupled electron spins in a
  room-temperature solid.
\newblock {\em Nat Phys}, 6(4):249--253, April 2010.

\bibitem{fuchs2008excited-state}
G.~D. Fuchs, V.~V. Dobrovitski, R.~Hanson, A.~Batra, C.~D. Weis, T.~Schenkel,
  and D.~D. Awschalom.
\newblock Excited-state spectroscopy using single-spin manipulation in diamond.
\newblock {\em Phys. Rev. Lett.}, 101:117601, 2008.

\bibitem{davies1976optical}
Gordon Davies and M~F Hamer.
\newblock Optical studies of the 1.945 {eV} vibronic band in diamond.
\newblock {\em Proc. R. Soc. Lond. A.}, 348:285--298, 1976.

\bibitem{loubser1977optical}
J.H.N. Loubser and J.A.~Van Wyk.
\newblock Optical spin-polarization in a triplet state in irradiated and
  annealed type 1b diamonds.
\newblock {\em Diamond Res.}, 1:11 -- 15, 1977.

\bibitem{loubser1978electron}
J.~H.~N. Loubser and J.~A.~van Wyk.
\newblock Electron spin resonance in the study of diamond.
\newblock {\em Reports on Progress in Physics}, 41(8):1201, August 1978.

\bibitem{reddy1987two-laser}
N.~R.~S. Reddy, N.~B. Manson, and E.~R. Krausz.
\newblock Two-laser spectral hole burning in a colour centre in diamond.
\newblock {\em J. Lumin.}, 38:46, December 1987.

\bibitem{oort1988optically}
E.~van Oort, N.B. Manson, and M.~Glasbeek.
\newblock Optically detected spin coherence of the diamond n-v centre in its
  triplet ground state.
\newblock {\em J. Phys. C}, 21:4385, 1988.

\bibitem{redman1991spin}
D.~A Redman, S.~Brown, R.~H Sands, and S.~C Rand.
\newblock Spin dynamics and electronic states of n-v centers in diamond by
  {EPR} and four-wave-mixing spectroscopy.
\newblock {\em Phys. Rev. Lett.}, 67(24):3420 -- 3423, December 1991.

\bibitem{manson2006nitrogen-vacancy}
N.~B. Manson, J.~P. Harrison, and M.~J. Sellars.
\newblock Nitrogen-vacancy center in diamond: Model of the electronic structure
  and associated dynamics.
\newblock {\em Phys. Rev. B}, 74(10):104303, 2006.

\bibitem{rogers2008infrared}
L.~J. Rogers, S.~Armstrong, M.~J. Sellars, and N.~B. Manson.
\newblock Infrared emission of the {NV} centre in diamond: Zeeman and uniaxial
  stress studies.
\newblock {\em New Journal of Physics}, 10(10):103024, 2008.

\bibitem{rogers2010how}
Lachlan Rogers.
\newblock How far into the infrared can a colour centre in diamond emit?
\newblock {\em Physics Procedia}, 3(4):1557--1561, February 2010.

\bibitem{goss1996twelve-line}
J.~P. Goss, R.~Jones, S.~J. Breuer, P.~R. Briddon, and S.~{\"O}berg.
\newblock The twelve-line 1.682 {eV} luminescence center in diamond and the
  vacancy-silicon complex.
\newblock {\em Physical Review Letters}, 77(14):3041--3044, September 1996.

\bibitem{manson2007issues}
N.B. Manson and R.L. McMurtrie.
\newblock Issues concerning the nitrogen-vacancy center in diamond.
\newblock {\em Journal of Luminescence}, 127(1):98--103, November 2007.

\bibitem{gali2008ab}
Adam Gali, Maria Fyta, and Efthimios Kaxiras.
\newblock Ab initio supercell calculations on nitrogen-vacancy center in
  diamond: Electronic structure and hyperfine tensors.
\newblock {\em Phys. Rev. B}, 77:155206, 2008.

\bibitem{delaney2010spin-polarization}
Paul Delaney, James~C. Greer, and J.~Andreas Larsson.
\newblock Spin-polarization mechanisms of the nitrogen-vacancy center in
  diamond.
\newblock {\em Nano Letters}, 10(2):610--614, February 2010.

\bibitem{kaplyanskii1964noncubic}
A~A Kaplyanskii.
\newblock Noncubic centers in cubic crystals and their piezospectroscopic
  investigation.
\newblock {\em Opt. Spectrosc.}, 16:329{\textendash}337, 1964.

\bibitem{kaplyanskii1964computation}
A~A Kaplyanskii.
\newblock Computation of deformation splitting of spectral transitions in cubic
  crystals.
\newblock {\em Opt. Spectrosc.}, 16:557{\textendash}565, 1964.

\bibitem{hughes1967uniaxial}
A~E Hughes and W~A Runciman.
\newblock Uniaxial stress splitting of doubly degenerate states of tetragonal
  and trigonal centres in cubic crystals.
\newblock {\em Proc. Phys. Soc.}, 90:827--838, 1967.

\bibitem{crowther1967phonon}
P.A. Crowther and P.J. Dean.
\newblock Phonon interactions, piezo-optical properties and the
  inter-relationship of the n3 and n9 absorption-emission systems in diamond.
\newblock {\em Journal of Physics and Chemistry of Solids}, 28(7):1115--1136,
  July 1967.

\bibitem{davies1979dynamic}
Gordon Davies.
\newblock Dynamic jahn-teller distortions at trigonal optical centres in
  diamond.
\newblock {\em J. Phys. C}, 12(13):2551--2566, 1979.

\bibitem{manson2010optically}
Neil Manson, Lachlan Rogers, Marcus Doherty, and Lloyd Hollenberg.
\newblock Optically induced spin polarisation of the {NV}- centre in diamond:
  role of electron-vibration interaction.
\newblock {arXiv} e-print 1011.2840, November 2010.

\bibitem{doherty2011negatively}
M~W Doherty, N~B Manson, P~Delaney, and L~C~L Hollenberg.
\newblock The negatively charged nitrogen-vacancy centre in diamond: the
  electronic solution.
\newblock {\em New Journal of Physics}, 13(2):025019, February 2011.

\bibitem{maze2011properties}
J~R Maze, A~Gali, E~Togan, Y~Chu, A~Trifonov, E~Kaxiras, and M~D Lukin.
\newblock Properties of nitrogen-vacancy centers in diamond: the group
  theoretic approach.
\newblock {\em New Journal of Physics}, 13(2):025025, February 2011.

\bibitem{ham1972jahn-teller}
Frank~S Ham.
\newblock Jahn-teller effects in electron paramagnetic resonance spectra.
\newblock In S.~Geschwind, editor, {\em Electron Paramagnetic Resonance}, pages
  1--119. Plenum Press, 1972.

\bibitem{coulson1957colour}
C.~A Coulson and Mary~J Kearsley.
\newblock Colour centres in irradiated diamonds. i.
\newblock {\em Proceedings of the Royal Society of London. Series A.
  Mathematical and Physical Sciences}, 241(1227):433--454, September 1957.

\bibitem{lenef1996electronic}
A.~Lenef and S.~C Rand.
\newblock Electronic structure of the n-v center in diamond: Theory.
\newblock {\em Phys. Rev. B}, 53(20):13441{\textendash}13455, May 1996.

\bibitem{manson2013assignment}
N.~B. Manson, K.~Beha, A.~Batalov, L.~J. Rogers, M.~W. Doherty,
  R.~Bratschitsch, and A.~Leitenstorfer.
\newblock Assignment of the {NV}{\textasciicircum}\{0\} 575-nm zero-phonon line
  in diamond to a {\textasciicircum}\{2\}e-{\textasciicircum}\{2\}a\_\{2\}
  transition.
\newblock {\em Physical Review B}, 87(15):155209, April 2013.

\bibitem{jones1997n2}
R.~Jones, J.~P. Goss, P.~R. Briddon, and S.~{\"O}berg.
\newblock N2 and \{N\}4 optical transitions in diamond: A breakdown of the
  vacancy model.
\newblock {\em Physical Review B}, 56(4):R1654--R1656, July 1997.

\end{thebibliography}

\end{document}